\documentclass[prl,twocolumn]{revtex4}
\usepackage{graphicx}
\renewcommand{\narrowtext}{\begin{multicols}{2} \global\columnwidth20.5pc}

\def\be{\begin{eqnarray}}
\def\ee{\end{eqnarray}}
\newcommand{\nn}{\nonumber\\}
\newcommand{\Eq}[1]{Eq.~(\ref{#1})}

\newcommand{\w}{{\omega}}
\newcommand{\cR}{{{\cal R}}}
\newcommand{\s}{{\sigma}}
\newcommand{\p}{{\partial}}

\begin{document}
\draft

\title{Stabilizing Superconductivity in Nanowires by Coupling to Dissipative Environments}
\author{Henry C. Fu$^{1}$, Alexander Seidel$^{2}$, John Clarke$^{1,2}$ and Dung-Hai Lee$^{1,2,3}$}
\affiliation{${1~}$Department of Physics,University of California,
Berkeley, CA 94720-7300, USA} \affiliation{${2}$ Materials
Sciences Division, Lawrence Berkeley National Laboratory,
Berkeley, CA 94720, USA} \affiliation{$3$ Center for Advanced
Study, Tsinghua University, Beijing 100084, China}

\date{\today}

\begin{abstract}
We present a theory for a finite-length superconducting nanowire
coupled to an environment. We show that in the absence of
dissipation quantum phase slips always destroy superconductivity,
even at zero temperature. Dissipation stabilizes the
superconducting phase. We apply this theory to explain the
``anti-proximity effect" recently seen by Tian {\it et. al.} in
Zinc nanowires.
\end{abstract}

\maketitle

The effect of dissipation on macroscopic quantum coherence is
currently a subject of considerable interest\cite{leggett}. In the
context of quantum bits ("qubits"), a dissipative environment always
increases the decoherence rate\cite{shoen}. On the other hand,
superconductivity in a Josephson junction is enhanced by
dissipation\cite{rsj}. In this paper we examine a particularly
striking example of the latter phenomenon: the stabilization of
superconductivity in a nanowire by dissipation in its environment.

In one-dimensional superconducting wires quantum phase
fluctuations can destroy long-range phase coherence even at zero
temperature; however, finite superfluid density can survive
through the Kosterlitz-Thouless (KT) physics\cite{KT}.  The
quantum action for a superconducting wire at zero temperature is
equivalent to that of the two-dimensional classical XY
model\cite{KT} at finite temperature. The two phases of the latter
correspond to the superconducting and insulating phases of the
wire. In particular KT vortices in the XY model correspond to
phase slip events in the wire in which the phase gradient unwinds
by $2 \pi$. The resistance of real nanowires can display both
insulating and superconducting behavior as temperature is
decreased\cite{newbower,giordano,expt}. Both thermal\cite{LAMH}
and quantum\cite{giordano, QPS} phase slips play an important role
in generating resistance and destroying superconductivity.
However, it is not settled what determines which nanowires are
superconducting\cite{expt}.  In particular, it is unclear whether
or not dissipation is important in determining the low-temperature
phase of nanowires\cite{buchler}.

In a recent experiment\cite{tian} Tian {\it et al.} observed an
unexpected effect when a 2-$\mu$m long, 40-nm diameter Zn nanowire
is sandwiched between two bulk superconducting electrodes. Under
zero applied magnetic field, when the electrodes are
superconducting, the Zn nanowire exhibits resistive behavior down to
the lowest measurement temperature. However after a sufficiently
strong magnetic field $B$ has suppressed the superconductivity of
the electrodes, the nanowire becomes superconducting at about 0.8 K.
Tian {\it et al.} dubbed this phenomenon the ``antiproximity effect"
(APE). Here we present a theory suggesting that this surprising
effect is due to the dissipation at the boundary between the
nanowire and electrodes. We show that when the nanowire has a finite
length, the ends of the wire are mapped onto two parallel boundary
lines that can screen the vortex-antivortex interaction in the XY
model. This screening destroys the superconducting phase even at
$T=0$. When the ends of the wire are coupled to a dissipative
environment, the screening becomes incomplete.  As a result, for
sufficiently large dissipation the superconducting phase is
stabilized.  The
importance 
of the boundary dissipation has been suggested by B\"uchler {\it
et al.}\cite{buchler}.

We begin by summarizing the experimental findings reported in
Ref.\cite{tian}.  Tian {\it et al.} prepared Zn nanowires in the
pores of porous polycarbonate or porous alumina membranes [Fig.
1(a)]. They pressed In or Sn wires on each side of the membrane to
form circular disks approximately 1 mm in diameter that made
contact to the ends of a single nanowire or, more generally, a
number of nanowires. By applying a magnetic field above the
critical field of the electrodes but below the critical field of
the Zn nanowire (which is enhanced by its small
diameter\cite{Tinkham}), they suppressed the superconductivity of
the electrodes. They measured the resistance and current-voltage
($I-V$) characteristics of their samples using the four-terminal
arrangement indicated in Fig. 1(a).
\begin{figure}
\includegraphics[width=8.5cm]{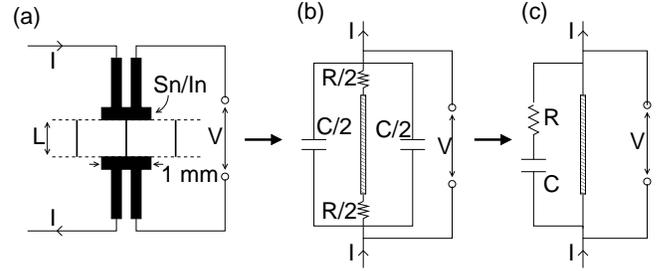}
\caption{Schematic diagrams of experiment by Tian {\it et al}. (a)
Experimental configuration: pores in polycarbonate membrane
contain nanowires, with In or Sn electrodes making contact to the
wire(s).  In this case, only one wire is contacted.  Current $I$ and
voltage $V$ are measured as shown.  (b) Circuit representation
showing contact resistances $R$/2 to each electrode and a
capacitance $C$/2 from each half of the parallel plate capacitor,
assuming the single wire is placed symmetrically.  (c) Simplified
model.  For the purpose of damping, $R$ and $C$ are connected in
series across the nanowire.}\label{diagram1}
\end{figure}
In this paper we focus on the sample Z4, the behavior of which is
shown in Fig. 3(b) of Ref.\cite{tian}. This sample had In
electrodes and is believed to have contained a single nanowire
with length $L$ = 2 $\mu$m.  When the In electrodes were driven
normal by the magnetic field, the Zn nanowire exhibited a
superconducting transition at a temperature that decreased as the
field was further increased.  In contrast, with zero applied field
(hence superconducting electrodes), the resistance showed a drop
of about 20 $\Omega$ as the temperature was lowered through the
transition temperature of the In (about 3.4 K), but the nanowire
{\it did not} go superconducting down to the lowest measurement
temperature (0.47 K).

Our model of the experiment is shown in Fig. 1(b). When the
electrodes are in the normal state, the nanowire is connected to
each of them via a contact resistance $R$/2.  When the electrodes
become superconducting, we assume that this resistance is
eliminated by the proximity effect; for sample Z4 [Fig. 3(b)] of
Ref.\cite{tian}, we estimate $R\approx 20 \Omega$   from the drop
in resistance when the electrodes become superconducting . We
estimate the resistance of the electrodes themselves in the normal
state to be on the order of $1$m$\Omega$ , which is negligible for
our present discussion.  The nanowire and its contact resistance
are in parallel with the capacitance of the parallel plate
capacitor formed by the two electrodes and the intervening
dielectric layer. This model can be simplified to a resistance $R$
and capacitance $C$ in series connected across the nanowire, as
depicted in Fig. 1(c).  This figure makes it clear that phase
fluctuations in the nanowire at frequencies above $f_0 = (2\pi RC)^{-1}$
induce currents through the shunting resistance and capacitor and are
thereby damped.  Using the area $A =\pi(0.5$mm$)^2$ of the
capacitor, the dielectric thickness L = $2\mu$m and assuming a
dielectric constant for polycarbonate  $\epsilon =
2.9$\cite{polyc}, we find $C =\epsilon\epsilon_0 A/L\approx 10$pF. 
Here, $\epsilon_0\approx 8.85\times 10^{-12} {\rm Fm}^{-1}$ is the
vacuum permittivity.

When the electrodes are superconducting the static resistance $R$
in Fig.1(c) vanishes. Thus for energy (frequency) less than the
bulk superconducting gap of the electrodes the quantum wire is
shunted by a capacitor. In the latter part of the paper we  show
that under such conditions the quantum fluctuations of the
superconducting phase drive the superfluid density of the zinc
nanowire to zero even at zero temperature. When the superfluid
density vanishes Cooper pairs disassociate and the wire becomes
normal. In that case the equivalent circuit of Fig.1(c)
becomes that of Fig.2 where the quantum wire acts as a normal
resistor with resistance $R_N$. This explains the ohmic behavior
at zero applied magnetic field, and the constant differential
resistance $dV/dI$ in the $I-V$ curve from zero voltage to about
 $500\mu$V.
\begin{figure}
\includegraphics[width=4.2cm]{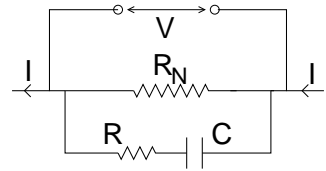}
\caption{Phase fluctuations ultimately drive the quantum wire
normal via depairing. The normal state resistance of the quantum
wire is $R_N$.}\label{diagram2}
\end{figure}

When the electrodes are driven normal by $B\ge 30$mT, the
boundary resistances between the Zn wire and the electrodes become
non-zero. Using the parameters $R\approx 20 \Omega$, $C\approx 10$pF 
we estimate $f_0 = (2\pi RC)^{-1}\approx
0.8 ~{\rm GHz}$. This gives $h f_0 \approx 0.04$K, an order of
magnitude lower than the thermal energy at the lowest measurement
temperature, $0.47$K. Thus in the relevant frequency/temperature
range the
external circuitry behaves as a pure resistor (i.e. $1/2\pi f
C\rightarrow 0$). As we show later, if this shunting resistance is
smaller than the quantum of resistance $h/4e^2\approx 6.4$k$\Omega$
it damps the superconducting phase fluctuations sufficiently to
stabilize superconductivity. If the measurement temperature is
lowered below $0.04$K we expect the residual quantum phase
fluctuations to destroy superconductivity and cause a reentrant
behavior. The fact that dissipation can stabilize
superconductivity is rather similar to the behavior of the
``resistively shunted (Josephson) junction'' (RSJ)\cite{rsj}.
However, unlike the RSJ, a quantum wire can undergo depairing when
the phase fluctuations are severe. As a result the normal state
resistance of the non-superconducting wire does not have to exceed
$h/4e^2$. Our theory is  consistent with the observation that the
APE practically vanished when one of the electrodes
was replaced with a non-superconducting metal.

In the remainder of the paper we present a theoretical analysis of
how dissipation suppresses phase slips in a superconducting wire.
Our main results are: (1) Through a duality transformation, we
establish the connection between  quantum phase slips and KT
vortices (instantons) in 1+1 dimensions. Our theory can be viewed
as an appropriate generalization of the RSJ model to quantum
wires. (2) At T=0 an isolated superconducting wire of finite
length is equivalent to a classical two dimensional electrostatic
problem where bulk charges (vortices) interact with two metallic
boundaries each representing the (imaginary-time) world-line
 of the endpoints of the quantum wire (Fig.3).
\begin{figure}
\includegraphics[width=4cm]{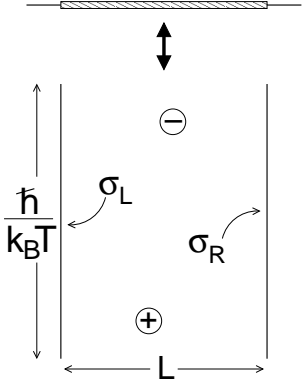}
\caption{A finite-length quantum wire is mapped onto a 1+1
dimensional Coulomb gas sandwiched between two metallic lines (the
world-line of the end points of the wire).  In the absence of
dissipation, the ``surface'' vortex density $\s_L$ and $\s_R$
completely screens the logarithmic interaction between bulk
vortices. As a result, vortex-antivortex pairs unbind and the wire
is normal. In the presence of dissipation, the logarithmic
interaction is not completely screened, and the
superconducting phase is stable for sufficiently small shunt
resistance.}\label{diagram3}
\end{figure}
Due to screening, vortices separated sufficiently far apart in the
imaginary time direction always unbind, and the wire is normal even
at $T=0$. (3) With shunt resistance $R$, the screening is
incomplete. For $R<h/4e^2$, and with $1/C=0$, the vortices remain
bound and the quantum wire is superconducting at $T=0$.


The imaginary-time action describing the quantum fluctuation of the
superconducting phase of a quantum wire is given by
$S=\int_{-\infty}^{\infty} dx\int_0^\beta dt{\cal{L}}$, where \be &&
{\cal{L}}=(K/2)|\partial_x\phi |^2+(1/2u)|\partial_t\phi
|^2.\label{qwire}\ee In \Eq{qwire} $\phi(x,t)=e^{i\theta(x,t)}$ is
the phase factor of the superconducting order parameter at position
$x$ and time $t$, $K$ is the superfluid density, and $u$ is the
inverse compressibility. At zero temperature
($\beta\rightarrow\infty$), depending on $K$ and $u$ there are two
possible phases: a superconducting phase and an insulating phase. In
the superconducting phase the topological singularity in $\phi$,
i.e. the vortices in space and time, are bound. In the insulating
phase, the space-time vortices (or instantons) proliferate. From
this point of view of the quantum wire a space-time vortex is a
quantum phase slip event.

In the presence of the external circuitry in Fig.1(c) the
action in \Eq{qwire} acquires an extra boundary term ($ S\rightarrow
S+S_{diss}$) with $S_{diss}=\int_0^\beta\int_0^\beta
dtdt'{\cal{L}}_{diss}$ given by 
\be
{\cal{L}}_{diss}=-(1/2)[\bar{\eta}(t)\partial_t\eta(t)]F(t-t')[\bar{\eta}(t')\partial_{t'}\eta(t')].\label{diss}\ee
In \Eq{diss} $\eta \equiv \phi(L/2,t)\bar{\phi}(-L/2,t)$ is the
relative phase factor of the two ends of the quantum wire
and \be
F(t-t')=(\pi/\beta\cR)\sum_{\w_n}|\w_n|^{-1}\exp[{i\w_n(t-t')}].\ee
Here $\cR$ is the dimensionless resistance $\cR=R/(h/4e^2)$, and
$\w_n=2\pi n/\beta$ is the Matsubara frequency. We omit the
capacitance of the external circuitry because we are interested in
the temperature range $k_BT>> hf_0\approx 0.04K$ where the
reactance of the capacitance is negligible. Clearly, this
assumption would be invalid if one were to repeat the experiment
of Tian {\it et al.}\cite{tian} at temperatures below $0.04K$; in
particular, this treatment can not predict the expected re-entrant
behavior.

Next we perform the standard duality transformation\cite{dhl}
keeping track of the finite spatial extent of the wire  to obtain
$S_D=\int_{-{L\over 2}}^{{L\over 2}} \! dx\int_0^\beta
\! dt \, {\cal{L}}_D+\int_0^\beta \! dt\int_0^\beta
\! dt' \, {\cal{L}}_{diss}+\int_0^\beta \! dt \, {\cal{L}}_{ends}$ where \be
&&{\cal{L}}_D={1\over 2K}[\p_t\chi(x,t)]^2+{u\over
2}[\p_x\chi(x,t)]^2-i\chi(x,t)\rho_v(x,t)\nn&&{\cal{L}}_{diss}={1\over
2}[\sigma_L(t)+\sigma_R(t)]~F(t-t')~[\sigma_L(t')+\sigma_R(t')]\nn&&{\cal{L}}_{ends}=
-i[\chi(L/2,t)\s_R(t)+\chi(-L/2,t)\s_L(t)].\label{vort}\ee In
\Eq{vort} $ \rho_v(x,t)=\sum_i Q_i\delta(x-x_i)\delta(t-t_i)$ is the
vortex density in space time, and \be\s_{R,L}(t)=\pm i
\bar{\phi}(\pm L/2,t)\p_t\phi(\pm
L/2,t)
.\ee After we
have integrated out $\chi$, the first term in \Eq{vort} is the
standard vortex Coulomb gas action, and the second term is due to
the coupling to the environment. The last term is a boundary term
arising from the finite
spatial extent of the quantum wire. 

Equation (\ref{vort}) is quadratic in $\chi$, so $\chi$ is 
integrated out exactly via the saddle point solution
$\delta S/\delta\chi=0$, yielding \be &&(1/K)\p_t^2\chi +
u\p_x^2\chi=-i\rho_v, ~u\p_x\chi\Big{|}_{\pm L/2}=\pm
i\sigma_{R,L}.~\label{sadd}\ee
If we define $\Phi(x,t)\equiv -i\chi(x,t)$ \Eq{sadd} becomes \be
&&(1/K)\p_t^2\Phi + u\p_x^2\Phi=-\rho_v,~u\p_x\Phi\Big{|}_{\pm
L/2}=\pm\sigma_{R,L}.\label{es}\ee Equation (\ref{es}) takes the
form of a 2D electrostatic problem with bulk charge density $\rho_v$
and surface charge density $\s_{L,R}$. 
In terms of $\Phi$, ${\cal{L}}_D$ and ${\cal{L}}_{ends}$ become
\be&&{\cal{L}}_D=-{1\over 2K}[\p_t\Phi(x,t)]^2-{u\over
2}[\p_x\Phi(x,t)]^2+\Phi(x,t)\rho_v(x,t)\nn&&{\cal{L}}_{ends}=
[\Phi(L/2,t)\s_R(t)+\Phi(-L/2,t)\s_L(t)].\label{esw}\ee

Substituting the solution $\Phi_c(x,t)$ of \Eq{es} into
Eqs.(\ref{esw}) and (\ref{vort}) we obtain an
action,$S_D[\s_L,\s_R,\rho_v]$, that depends only on $\rho_v$ and
$\s_{L,R}$: \be &&S_D={1\over 2}\int_{-{L\over 2}}^{L\over
2}dx\int_0^\beta dt\Phi_c(x,t)\rho_v(x,t)+\int_0^\beta
dt\int_0^\beta dt' {\cal{L}}_{diss}\nn&& +\int_0^\beta dt
[\Phi_c(L/2,t)\s_R(t)+\Phi_c(-L/2,t)\s_L(t)].\label{vort1}\ee To
obtain a final action that depends only on the vortex density
$\rho_v$ we integrate out $\s_R,\s_L$.  Since \Eq{vort1} depends
on $\s_R,\s_L$ only quadratically we can again use the saddle
point method, solving $\delta S/\delta\s_{L,R}=0$ and substituting
the solution back into \Eq{vort1}.  In a wire of length $L$, the
result is that vortices with spatial coordinates $x_i$ ($-L/2\le
x_i\le L/2$) interact via \be S_D&=&\sum_{i\ne j} Q_i
G(x_i,x_j;t_i-t_j) Q_j. \label{SD}\ee In \Eq{SD}
$G(x_i,x_j;t_{ij}) = \frac{1}{\beta} \sum_{\omega_n} (G_1 + G_2)
e^{i\w_n t_{ij}}$ with \be G_1 &=& \sqrt{\frac{K}{u}} \frac{
c_{L-|x_i-x_j|} - c_{x_i + x_j}}{\omega_n s_L}\nn G_2 &=&
\sqrt{\frac{K}{u}} \frac{ c_{x_i} c_{x_j}}{\omega_n s_L} \left( 1
+
  \frac{sgn(\omega_n) \cR s_L}{4 \pi s_{L/2}^2} \right)^{-1}.\label{exact2}
\end{eqnarray}
Here $c_x =\cosh(\omega_n x/\sqrt{Ku})$ and $s_x =\sinh(\omega_n
x/\sqrt{Ku})$. Expanding Eq.(\ref{exact2}) for small
$\omega_n$ shows that vortices with time separation much
greater than $L/\sqrt{uK}$ interact via \be S_D=\sum_{i\ne j} Q_i
Q_j \Big[-{(1/2\cR)}\ln{(|t_{ij}|/
\tau_0)}+J_{ij}(t_{ij})\Big].\label{final}\ee In \Eq{final}
$J_{ij}$ is a short-range interaction (in time).  It is important
to note that the long-range logarithmic interaction is controlled
only by the resistive dissipation.
Aside from the irrelevant short-range interaction $J_{ij}$,
\Eq{final} is identical to the phase slip action of a single RSJ
(we identify $Q_j$ with the phase slip). For $\cR<1$ the phase
slip and anti-phase slip form bound pairs, while for $\cR>1$ the
phase slip and anti-phase slip unbind\cite{rsj,review}.  The
former corresponds to the superconducting phase of the quantum
wire, while the latter corresponds to the non-superconducting
phase.

For an isolated quantum wire a similar calculation leads to a
short-range action for the vortices [\Eq{SD}] with
$G(x_i,x_j;t_i-t_j) = \frac{1}{\beta} \sum_{\omega_n} G_1
e^{i\w_n(t_i-t_j)}$.
Since the interaction is short-ranged, phase slips and anti-phase
slips always unbind. As a result a free, finite-length quantum
wire is always non-superconducting, even at zero temperature. Our
results are fully consistent and agree in spirit with those of
B\"uchler {\it et al.}\cite{buchler}, who derived the phase slip
interaction from phenomenological boundary conditions. In the
current work, we derive the boundary effects and the phase slip
interaction in the presence of dissipation {\it exactly}.
Consequently we have an explicit phase slip interaction valid at
both long and short time scales as  is required for a quantitative
understanding of the phase slip physics of a quantum wire.

As emphasized earlier, a piece of physics of the quantum wire not
present in the RSJ is the fact that as phase fluctuations suppress
the superfluid density to zero, Cooper pair disassociation will
take place before the wire becomes a Cooper pair insulator. Once
electrons depair, the wire can no longer be described by the
phase-only action given in this paper. This is quite similar to
the case of superconductor to non-superconductor quantum phase
transition of homogeneous films, where electron tunneling always
finds a closing energy gap at the phase transition\cite{dyne}.
Thus, the normal state resistance of the wire is not directly
related to the shunt resistance $R$ used in our purely bosonic
model and which determines the fate of the wire. Finally, we note
that the asymptotic interaction between vortices far separated in
time is valid when $\hbar/k_BT >> L/\sqrt{uK}$. For non-zero
temperatures, a sufficiently long nanowire acts as an infinite
wire. When the KT vortices are bound, the nanowire exhibits the
transport properties of an attractive Luttinger liquid\cite{lutt}
and hence exhibits superconducting-like characteristics.

To conclude, in this paper we present a theory for the quantum phase
slips of a superconducting nanowire, and the effect of environmental
dissipation on them. We apply this theory to explain the
anti-proximity effect recently observed by Tian {\it et al}. We
attribute the recurrence of superconductivity when the electrodes
are driven normal by a magnetic field to the onset of dissipation
from the boundary resistance between the quantum wire and the
electrodes. This dissipation suppresses  phase
fluctuation in the wire and stabilizes superconductivity.  

{\bf Acknowledgement}: We thank MingLiang Tian and Moses Chan for
providing very useful information on their measurements. We also
thank Z-Y Weng, X-G Wen, and  A. Viswanath for useful discussions.
This work was supported by the Directior, Office of Science,
Office of Basic Energy Sciences, Materials Sciences and
Engineering Division, of the U.S. Department of Energy under
Contract No. DE-AC02-05CH11231 (AS, JC and DHL).


\end{document}